\documentclass[10pt,english,letterpaper,twoside,conference]{IEEEtran}

\usepackage[T1]{fontenc}
\usepackage[latin9]{inputenc}
\usepackage{amsmath}
\usepackage{amssymb}
\usepackage{psfrag}
\usepackage[dvips]{graphicx}

\makeatletter
\newtheorem{thm}{Theorem}
\newtheorem{proposition}{Proposition}
\newtheorem{corollary}{Corollary}
\newtheorem{example}{Example}

\makeatletter



\makeatother

\usepackage{babel}
\makeatother

\begin{document}

\title{An Entropic View of Pickands' Theorem}

\author{ \authorblockN{Jean-Fran\c cois Bercher} \authorblockA{Laboratoire
des Signaux et Syst\`emes, \\
 CNRS-Univ Paris Sud-Supelec, \\
 91192 Gif-sur-Yvette cedex, France\\
 Email: bercherj@esiee.fr}
 \and
 \authorblockN{Christophe Vignat}
\authorblockA{Institut Gaspard Monge\\
Universit\'e de Marne la Vall\'{e}e \\
77454 Marne-la-Vall\'ee cedex 02, France \\
 Email: vignat@univ-mlv.fr } }

\maketitle

\section*{Abstract}

It is shown that distributions arising in R\'enyi-Tsallis maximum entropy
setting are related to the Generalized Pareto Distributions (GPD)
that are widely used for modeling the tails of distributions. The
relevance of such modelization, as well as the ubiquity of GPD in
practical situations follows from Balkema-De Haan-Pickands theorem
on the distribution of excesses (over a high threshold). We provide
an entropic view of this result, by showing that the distribution
of a suitably normalized excess variable converges to the solution
of a maximum Tsallis entropy, which is the GPD.
This result resembles the entropic approach to the Central Limit theorem as provided in  \cite{barron}; however, the  convergence in entropy proved here is weaker than the convergence in supremum norm given by Pickands' theorem. 

\bigskip{}




\section{Introduction}

Generalized Pareto Distributions (GPD) are widely used in practice
for modeling the tails of distributions. The underlying rationale
is the Balkema-De Haan-Pickands theorem \cite{pickands_statistical_1975,balkema_residual_1974},
which asserts that the distribution function of the excess variable
$X-u|X>u$ (i.e. the distribution of the shifted variable $X$
exceeding a threshold $u$) converges, as $u \rightarrow \infty$, to a GPD with survival function:
\begin{equation}
S_{X}(x)=Pr\left(X>x\right)=\left(1+\frac{\gamma}{\sigma}x\right)^{-\frac{1}{\gamma}},
 \label{eq:defGPD}
\end{equation}
 where $\sigma$ is a scale parameter and $\gamma$ a shape parameter; for $\gamma=0,$ the GPD reduces to the exponential distribution $S_{X}(x)=\exp{(-x/\sigma)}$.
The corresponding density is 
\[
 f_X(x)=\frac{1}{\sigma}\left(1+\frac{\gamma}{\sigma}x\right)^{-\frac{1}{\gamma}-1},
\]
for $\gamma\neq 0$, and reduces to $ f_X(x)={1}/{\sigma} \exp(-x/\sigma)$ for $\gamma=0$.

In applied fields, GPD have encountered a large success since they
were obtained as the maximizers of a special entropy, the Tsallis
(Havrda-Charv\'{a}t-Dar\'ovczy) entropy  \cite{tsallis_possible_1988},
with suitable constraints. This entropy is defined by
\begin{equation}
H_{q}(f_X)=\frac{1}{1-q}\left(\int f_X^{q}\left(x\right) dx - 1 \right)
\label{eq:entropytsallis}
\end{equation}
for $q \geq 0$. We note that Shannon entropy 
\begin{equation}
H_{1}(f_X)=\lim_{q\rightarrow 1} H_q = -\int f_X\left(x\right) \log f_X(x)  dx
\label{eq:entropyshannon} 
\end{equation}
is recovered in the limit case $q=1$. 

It is worth mentioning that any monotonous
transform of the latter entropy exhibits the same GPD maximizers:
an important example is R\'enyi entropy \cite{Renyi1961}. 
The GPD distribution is of very high interest in many physical systems, since
it enables to model power-law phenomena. Indeed, power-laws are especially
interesting since they appear widely in physics, biology, economy,
and many other fields \cite{newman_power_2005}.

%
%
%
%
\begin{figure}

 \begin{center}
 \psfrag{xx}{$x$} 
\psfrag{g0}{\small $\gamma=0$}
  \psfrag{g1}{\small $\gamma=1$}
  \psfrag{g10}{\small $\gamma=10$}
  \psfrag{g100}{\small $\gamma=100$}
 \psfrag{formuleGPD}{$f_X(x)=\frac{1}{\sigma}\left(1+\frac{\gamma}{\sigma}x\right)^{-\frac{1}{\gamma}-1}$}
 \includegraphics[scale=0.33]{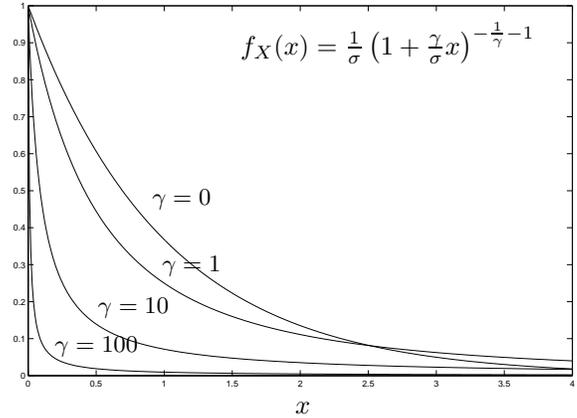}
\caption{Infinite support Generalized Pareto Densities for several values of the parameter $\gamma$, with $\sigma=1$. \label{fig:infinite-support-q-exponential}}

\end{center}
\end{figure}
\

In this communication, we give an interpretation of Pickands' 
theorem which relates it to the maximum (R\'enyi/Tsallis) entropy setting;
this view gives a possible interpretation for the
ubiquity of `Tsallis' (GPD) distributions in physics applications,
as well as in other fields, an an argument in support to the use of
R\'enyi/Tsallis entropies.

\

In the following, we deal with univariate distributions defined on
$\mathbb{R}$ or on a subset of $\mathbb{R}$. Our approach is as
follows: first, we show that the GPD can be obtained as the solution of
a  maximum
R\'enyi-Tsallis entropy problem with proper normalization and moment constraints. 
Second,
we consider distributions in the Fr\'echet domain of attraction of  distributions: this family includes for instance Cauchy, Student and Pareto distributions.
We characterize the associated $q$-norm and first moment of the survival
function associated to the excess variable $X-u|X>u$. Using an appropriate
normalization, we define a variable whose survival function's $q$-norm
and moment converge to constant values. 
We perform the same analysis for a subset of distributions in the domain of attraction of the Gumbel distributions.
Third, we show that
the distribution of excesses 
coincides asymptotically with the maximum Tsallis entropy solution. 

\section{Solution to the maximization of Tsallis' entropy}

We first 
derive
the expression of the solution to the maximization
of Tsallis' entropy subject to normalization and moment constraints.

\begin{proposition}
\label{prop1}
Consider the set ${\cal \mathcal{F}}=\left\{ G : \mathbb{R}^{+}\rightarrow\mathbb{R}\right\} .$
The maximum Tsallis entropy problem (or equivalently the maximum $q$-norm
problem), with $q < 1$, defined by
\begin{equation}
\left\{
\begin{aligned}
& \max_{G\in\mathcal{F}} H_{q}(G) 
\\ 
& \text{ subject to  } 
\intop_{0}^{+\infty} zG(z)dz=\mu\mathrm{\: and\:}\intop_{0}^{+\infty} G(z)dz=\theta\label{eq:MEConstraints}
\end{aligned}
\right.
\end{equation}
 has for unique solution \begin{equation}
G_{*}(z)=\alpha^{\frac{1}{q-1}}\left(1+\frac{\beta}{\alpha}z\right)^{\frac{1}{q-1}} \text{for } q\neq 1  \label{eq:optME}
\end{equation}
\def\thefootnote{\fnsymbol{footnote}}
where  $\alpha\ge0$ and $\beta\ge0$. Moreover, for $1/2<q< 1$\begin{footnote}{Note that the mean is not defined for $q<1/2$} \end{footnote}

\begin{equation}
\mu=\frac{\left(q-1\right)^{2}}{q\left(2q-1\right)}\frac{\alpha^{\frac{2q-1}{q-1}}}{\beta^{2}},\:\:
\theta=\frac{\alpha^{\frac{q}{q-1}}}{\beta}\frac{\left(1-q\right)}{q}\label{eq:ValuesOptConstraint}\end{equation}
 \begin{equation}
\mathrm{and\:\:}||G_{*}||_q^q=\frac{\alpha^{\frac{2q-1}{q-1}}}{\beta}\frac{\left(1-q\right)}{\left(2q-1\right)}.{\normalcolor }\label{eq:ValuesOptEntropy}\end{equation}
In the case $q=1,$ the unique solution writes
and 
\begin{equation}
G_{*}(z)=\alpha \exp(-\beta x) 
\label{eq:optMEShannon}
\end{equation}
with constants $\alpha$ and $\beta$ such that
\begin{equation}
\mu = \frac{\alpha}{\beta^2}, \,\, \theta = \frac{\alpha}{\beta}
\end{equation}
and the Shannon entropy is
\[
H_{1}(G_{*})=-\frac{\alpha}{\beta}\log{\alpha} + \alpha
\]

\end{proposition}

\begin{proof}
The solution of the maximum Shannon ($q=1$) entropy problem is well documented. We only consider here the $q < 1$ case and we follow the approach of \cite{vignat_about_2004}. Consider
the functional Bregman divergence: \begin{equation}
\begin{split}
B_{}(f,g) & =\int d(f,g)dx 
\\ & =-\int\left(f(x)^{q}-g(x)^{q}\right)g(x)^{q-1} dx \\
& + {q} \int \left(f(x)-g(x)\right)g(x)^{q-1} dx
\end{split}
\end{equation}
 associated to the (pointwise) Bregman divergence $d(f,g)$ built
upon the strictly convex function $-x^{q}$ for $q\in(0,1)$. 
Then
let us evaluate the divergence between the distribution $G_{*}(z)$
in (\ref{eq:optME}) and any distribution $G(z),$ with $G$ dominated by $G_*$, $G(z)\ll G_{*}(z)$, and
satisfying (\ref{eq:MEConstraints}): 
\begin{align}
B(G,G_{*}) & =-\int_{\mathcal{S}} \left(G(z)^{q}-G_{*}(z)^{q}\right) dz \nonumber \\
&-{\alpha}\int_{\mathcal{S}} (G(z)G_{*}(z)^{q-1}-G_{*}(z)^{q})dz \nonumber \\
 & =-\int_{\mathcal{S}}G(z)^{q}dz+\int_{\mathcal{S}}{G_{*}}(z)^{q}dz,\label{eq:Dpos}
\end{align}
 where $\mathcal{S}$ denotes the support of $G_{*}(z)$. The last
line follows from the fact that since $G$ and $G_{*}$ both satisfy
(\ref{eq:MEConstraints}), then, using (\ref{eq:optME}) it is easy
to check that \[
\int_{\mathcal{S}}G(x)G_{*}(x)^{q-1}dx=\int_{\mathcal{S}}G_{*}(x)^{q}dx.\]
 The Bregman divergence $B(G,G_{*})$ being always positive and equal
to zero if and only if $G=G_{*}$, the equality (\ref{eq:Dpos}) implies
that, for $q\in[0,1[$, \begin{equation}
H_{q}(G_{*})\geq H_{q}(G)\end{equation}
which means that $G_{*}$ is the distribution with maximum R\'enyi-Tsallis
entropy, with $q\in[0,1[,$ in the set of all distributions $G\ll G_{*}$
satisfying the constraints (\ref{eq:MEConstraints}). Values of the
constraints (\ref{eq:ValuesOptConstraint}) and of the maximum entropy
(\ref{eq:ValuesOptEntropy}) follow by direct calculation. 
\end{proof}

\section{The distribution of excesses for distributions in Fr\'echet domain of attraction}

In the following, we consider 
the Fr\'echet domain
of attraction: this is the set $\mathcal{F}$ of distributions  such that if
variables $X_i$ are independent and identically distributed according to 
one of them, then $\max_{i=1..n} \{X_i\}$ converges to the 
GPD distribution as $n \to \infty$. 
It was shown by Gnedenko \cite{gnedenko_1943}
that a necessary and sufficient condition for a distribution to be in the Fr\'echet domain
of attraction is that its survival function $S(z)$ satisfies
\[
\lim_{z\rightarrow+\infty}\frac{S(z)}{S(cz)}=c^{a},\]
 for all $c>0$ and for some tail index $a>0$. Equivalently, this reads 
 \[
S(z) = z^{-a}l(z),
\]
 where $l(z)$ is a slowly varying function, i.e. a function such that $\lim_{z\rightarrow+\infty}\frac{l\left(zt\right)}{l\left(z\right)}=1,\,\,\forall t>0.$

Let us consider the excess variable $X_{u}=X-u|X>u$. Its survival
function is \[
S_{X_{u}}\left(z\right)=\frac{S_{X}\left(z+u\right)}{S_{X}\left(u\right)}.\]

\begin{proposition}
\label{theo:SXuFrechet}
Suppose that 
$X$ belongs to the Fr\'echet domain, with \[
S_{X}\left(z\right)\sim z^{-a}l\left(z\right),\]
 then $S_{X_{u}}$ has asymptotic
 $q-$norm 
 \[
\Vert S_{X_{u}}\Vert_{q}\sim 
( \frac{u}{aq-1} )^{1/q}
\]
 and asymptotic first moment, with $a>2$, \[
\int_{0}^{+\infty}zS_{X_{u}}\left(z\right)dz=\frac{u^{2}}{\left(1-a\right)\left(2-a\right)}.\]
\end{proposition}

\begin{proof}
the $q-$power of the $q-$norm writes
\begin{multline}
\Vert S_{X_{u}}\Vert_{q}^{q} =  \int_{0}^{+\infty}\left(\frac{S_{X}\left(z+u\right)}{S_{X}\left(u\right)}\right)^{q}dz \\
 =  \int_{u}^{+\infty}\left(\frac{S_{X}\left(z\right)}{S_{X}\left(u\right)}\right)^{q}dz 
 =  u\int_{1}^{+\infty}\left(\frac{S_{X}\left(wu\right)}{S_{X}\left(u\right)}\right)^{q}dw\\
\sim  u\int_{1}^{+\infty}\frac{\left(uw\right)^{-aq}}{u^{-a}}dw 
  =  u\int_{1}^{+\infty}w^{-aq}dw=\frac{u}{aq-1}, \nonumber
\end{multline}
with $1-aq\leq0,$ since $a>2,q>1/2$. Of course, we immediately
obtain, taking $q=1$, that
\[
\Vert S_{X_{u}}\Vert_{1}^{}=\frac{u}{a-1}.\]
Similarly, the first moment is
\begin{align*}
\int_{0}^{+\infty}zS_{X_{u}}\left(z\right)dz  & =  \int_{0}^{+\infty}z\frac{S_{X}\left(z+u\right)}{S_{X}\left(u\right)}dz \\
  =  \int_{u}^{+\infty}\left(z-u\right)\frac{S_{X}\left(z\right)}{S_{X}\left(u\right)}dz 
 & =  \int_{1}^{+\infty}u\left(w-1\right)\frac{S_{X}\left(wu\right)}{S_{X}\left(u\right)}udw\\ 
  \sim  u^{2}\int_{1}^{+\infty}\left(w-1\right)w^{-a}dw 
  & = \frac{u^{2}}{\left(1-a\right)\left(2-a\right)}.  \end{align*}
\end{proof}

We have a simple corollary to this result:
\begin{corollary}
\label{cor1}
The survival function $S_{Y_{u}}$ of random variable
$Y=X/g\left(u\right),$ where function {$g$}
is such that $g\left(u\right)\sim u$, has asymptotic norms
\[
\Vert S_{Y_{u}}\Vert_{q}\sim (\frac{1}{aq-1})^{1/q}\:\:\mathrm{and}\:\;\Vert S_{Y_{u}}\Vert_{1}^{}=\frac{1}{a-1}.\]
 and an asymptotic first moment \[
\int_{0}^{+\infty}zS_{Y_{u}}\left(z\right)dz\sim\frac{1}{\left(1-a\right)\left(2-a\right)}.\]
\end{corollary}

\begin{proof}
 The results for $S_{Y_{u}}$ follow directly from Proposition~\ref{theo:SXuFrechet}, with
 \begin{equation}
S_{Y_{u}}\left(z\right)=S_{X_{u}}\left(zg\left(u\right)\right).\label{eq:SurvivalYu}\end{equation}

\end{proof}

\section{The distribution of excesses for a subset of distributions in Gumbel domain of attraction}

For distributions in the Gumbel domain of attraction, the maximum of a 
set of variables converges to the Gumbel extreme value distribution. These distributions are characterized by an ``exponential'' fall-off, and are said ``light tailed'' distributions. Their excesses over a threshold are exponentially distributed, which corresponds to a GPD with $\gamma=0$. 
The general characterization of the Gumbel domain of attraction involves an inconvenient condition on the derivative of the hazard function. We 
consider here only the Weibull subset $\mathcal{W}$ of the Gumbel domain of attraction, whose survival functions verify
\begin{equation}
S(z) \sim \exp\left(- z^\xi l(z) \right)  
\end{equation}
where $\xi$ is the tail index and $l(z)$ is a slowly varying function. This set contains for example the Gaussian and Gamma distributions: 
\def\erfc{\mathrm{~erfc}}
\begin{itemize}
\item
The  survival function of a Gaussian distribution with zero mean and unitary variance is ${S}_X(x)=1/2 \erfc(x/\sqrt{2})$, with $\erfc$ the complementary error function. For $x\rightarrow +\infty$, the complementary error function
is equivalent to 
$\erfc(x)\sim \exp(-x^2)/(x\sqrt{\pi})$, and
\[
 S_X(x)=\frac{1}{\sqrt{2\pi}} \frac{e^{-\frac{x^2}{2} }}{x}=\frac{1}{\sqrt{2\pi}} e^{-\frac{x^2}{2} \left(1+2\frac{\log(x)}{x^2}\right) }
\]
%
\item
The survival function of a Gamma distribution with shape parameter $a$ and rate parameter $b$ is given by  ${S}_X(x)= \Gamma(a,bx)/\Gamma(a)$. Since $\Gamma(a,bx)\sim (bx)^{a-1}e^{-bx}$, we obtain
\[
 S_X(x)=\frac{1}{\Gamma(a)} e^{-bx\left( 1-\frac{(a-1)\log(bx)}{bx}\right)}
\]
%
\end{itemize}

\begin{proposition}
\label{theo:SXuGumbel}
Suppose that 
$X$ belongs to $\mathcal{W}$, with 
\[
S_{X}\left(z\right)\sim \exp(-z^{\xi}l\left(z\right)),
\]
 then the survival function $S_{X_{u}}$  of the excesses of $X$  has asymptotic
 Shannon entropy
 \[
H_1(S_{X_{u}})\sim 
 \frac{u^{1-\xi}}{\xi l(u)} 
\]
and asymptotic first moment and $1-$norm
\[
\int_{0}^{+\infty}zS_{X_{u}}\left(z\right)dz \sim
\frac{u^{2(1-\xi)}}{\xi^2 l(u)^2}, 
\,\,\,
\int_{0}^{+\infty}S_{X_{u}}\left(z\right)dz \sim
 \frac{u^{1-\xi}}{\xi l(u)} 
\]\\
\end{proposition}

\begin{proof}
The computations are essentially the same as 
in the proof of Proposition \ref{theo:SXuFrechet}.\\
\end{proof}

From this result, we deduce the

\begin{corollary}
\label{cor2}
The survival function $S_{Y_{u}}$ of random variable
$Y=u^{\xi -1} l(u) X$ has asymptotic Shannon entropy
\[
H_1(S_{Y_{u}})\sim \frac{1}{\xi},
\]
and asymptotic first moment and $1-$norm 
\[
\int_{0}^{+\infty}zS_{Y_{u}}\left(z\right)dz \sim
\frac{1}{\xi^2},
\,\,
\int_{0}^{+\infty}S_{Y_{u}}\left(z\right)dz \sim
 \frac{1}{\xi} 
\]
\end{corollary}

\section{The entropy solution and the distribution of excesses}

Let us now show that the distributions of excesses, both in the Gumbel and the Fr\'echet case, coincide with the maximum Tsallis entropy solution. 


In the Fr\'echet domain of attraction, this result reads as follows

\begin{thm}
\label{theo:frechetME}
if $X$ belongs to the Fr\'echet domain of attraction, then choosing $q<1$ such that
\[
a=\frac{1}{1-q},\]
the distribution of the excesses $Y_{u}$ as defined in Corollary \ref{cor1} reaches asymptotically the maximum
$q-$norm solution under constraints asymptotically equal to $\mu$
and $\theta$ provided $\alpha=\beta=1.$ 
\end{thm}
\begin{proof}
Choosing $a=\frac{1}{1-q}$ yields\[
\Vert S_{Y_{u}}\Vert_{q}^q \sim\frac{1}{aq-1}=\frac{1-q}{2q-1},\]
 \[
\Vert S_{Y_{u}}\Vert_{1}\sim\frac{1}{a-1}=\frac{1-q}{q}\]
and
\[
 \int_{0}^{+\infty}zS_{Y_{u}}\left(z\right)dz\sim\frac{1}{\left(1-a\right)\left(2-a\right)}=\frac{\left(q-1\right)^{2}}{q\left(2q-1\right)}\]
which coincide with the unique maximum $q-$norm function with constraints
$\mu$ and $\theta$ if and only if $\alpha=\beta=1.$ 

Since the maximum entropy solution with the same constraints  is unique, 
we obtain that the excess variable from a distribution
in the domain of attraction of Fr\'echet distribution asymptotically
follows a Generalized Pareto Distribution.
\end{proof}

~\\
In the Gumbel case, we obtain similarly

\begin{thm}
\label{theo:gumbelME}
if $X$ belongs to $\mathcal{W}$ with
\[
S_{X}(z)\sim \exp(-z^{\xi}l(z))
\]
then the distribution of the excesses $Y_{u}$ as defined in Corollary \ref{cor2} reaches asymptotically the maximum
Shannon entropy solution under constraints asymptotically equal to $\mu$
and $\theta$ provided $\alpha=1$ and $\beta=\xi.$ 
\end{thm}

\begin{proof}
Equating the Shannon entropy and the first moment and $1-$norm of the maximum entropy solution of Proposition  \ref{prop1} with the same quantities reached asymptotically by $S_{Y_{u}}$ as in Corollary \ref{cor2} yields $\alpha=1$  and $\beta=\xi$.
\end{proof}


~\\
\begin{example}
 As an illustration, let us consider the Cauchy case. The pdf is given by
\[
f_X\left(x\right)=\frac{2}{\pi\left(1+x^{2}\right)},\,\, x\ge0.
\]
Its survival function is $S_X(x)=1-\frac{2}{\pi}\arctan{(x)}$. Using now the fact that $\arctan(x)\approx \pi/2 -1/x$, 
for $x\gg 1$, we obtain that $S_X(x)\sim{2}/{(\pi x)}$, which means that the Cauchy distribution is in the Fr\'echet domain of attraction, with exponent $a=1$.

The survival function of the excess variable $X_u$ writes
\[
 {S}_{X_u}=\frac{{S}_X(x+u)}{{S}_X(u)}=\frac{1-\frac{2}{\pi}\arctan{(x+u)} }{1-\frac{2}{\pi}\arctan{(u)} }.
\]
Using the $\arctan(x)$ approximation again, we readily obtain, with the threshold $u \gg 1$,
\[
  {S}_{X_u} \sim \frac{1}{u+x} / \frac{1}{u} = \left(1+\frac{x}{u}\right)^{-1}
\]
which has the form of the Generalized Pareto Distribution (\ref{eq:defGPD}) with index $\gamma=1$. 
Finally, with $S_{Y_u}(x)=S_{X_u}(ux)$, we have $S_{Y_u}(x)\sim1/(1+x)$. \\

%
%

\end{example}

Since we want to emphasize on the resemblance with the Central limit theorem, we mention the following stability property of the  GPD (\ref{eq:defGPD}): the distribution of the excesses
over a threshold of GPD remains a GPD, with the same exponent but
a different shape parameter. 
This property is to be compared with the usual stability by addition of independent Gaussian random variables.

\begin{thm}
Given a GPD with parameters $\gamma,$ $\sigma$ the distribution
of excesses remains a GPD with parameters $\gamma$ and $\sigma'=\sigma\left(1+\frac{\gamma}{\sigma}u\right).$ 
\end{thm}
\begin{proof}
As usual, let $u$ denotes the threshold, $S_{X}$ the survival function
of the original GPD and $S_{X_{u}}$ the survival function of variable
$X_{u}.$Then, \[
S_{X_{u}}(x)=\frac{S_{X}(x+u)}{S_{X}(u)}=\frac{\left(1+\frac{\gamma}{\sigma}\left(x+u\right)\right)^{-\frac{1}{\gamma}}}{\left(1+\frac{\gamma}{\sigma}u\right)^{-\frac{1}{\gamma}}}=\left(1+\frac{\gamma}{\sigma'}x\right)^{-\frac{1}{\gamma}}\]
with $\sigma'=\sigma\left(1+\frac{\gamma}{\sigma}u\right).$
\end{proof}
Note that in the limit $\gamma=0$ case, the exponential distribution is invariant by thresholding, i.e. $\sigma'=\sigma.$

\section{Final comments}

In the Fr\'echet domain of attraction as well as in a subset of the Gumbel domain of attraction, we have connected the solution of a maximum $q$-entropy (or maximum $q$-norm) problem with the asymptotic distribution of excesses over a threshold, and showed that these distributions are Generalized Pareto Distributions. 
With this result, it is possible to connect the ubiquity of heavy-tailed distributions in physics, economics or signal processing, the distribution of the excesses over a threshold, and a maximum entropy construction. 

Our approach shows the convergence {\it in  entropy} of the distributions of excess over a threshold; this type of convergence is in fact weaker than the distribution in supremum norm proved in Pickands' theorem. However, this work underlines an interesting parallel with the entropic proof of the Central Limit Theorem as given in \cite{barron}.


\bibliographystyle{IEEEtran} 
\bibliography{fluctME,gpdpap}
\enlargethispage{0.5cm}

\end{document}